\documentclass{article}%
\usepackage{amsfonts}
\usepackage{amsmath}
\usepackage{amssymb}
\usepackage{graphicx}%
\newtheorem{theorem}{Theorem}

\newtheorem{remark}[theorem]{Remark}

\begin{document}

\title{Remarks on the Fact that the Uncertainty Principle Does Not Determine the
Quantum State}
\author{Maurice de Gosson, Franz Luef\\Faculty of Mathematics, University of Vienna }
\maketitle

\begin{abstract}
We discuss the relation between density matrices and the uncertainty
principle; this allows us to justify and explain a recent statement by Man'ko
\textit{et al}. We thereafter use Hardy's uncertainty principle to prove a new
result for Wigner distributions dominated by a Gaussian and we relate this
result to the coarse-graining of phase-space by \textquotedblleft quantum
blobs\textquotedblright.\ 

\end{abstract}

\section{Introduction}

Identifying the class of all phase-space functions who are Wigner
distributions of some mixed quantum state is a formidable and unfinished task.
The problem is actually the following: assume that a function $W(x,p)$ on
phase space defines, via the Weyl correspondence, a self-adjoint operator
$\widehat{\rho}$ with unit trace. Then $\widehat{\rho}$ is, a priori, an
excellent candidate for being the density operator of some mixed state with
Wigner distribution $W(x,p)$, provided that in addition this operator is
\textit{positive} (that is $\langle\widehat{\rho}\psi\left\vert \psi
\right\rangle \geq0$ for all square integrable $\psi$). And this is where the
difficulty comes from: outside a few well-known cases (for instance when
$\rho$ is a Gaussian), it is notoriously difficult in general to check the
positivity of $\widehat{\rho}$ by simply inspecting the function $W(x,p)$ (the
condition $W(x,p)\geq0$ is neither necessary nor sufficient, in strong
opposition to the classical case). Although there are general (and difficult)
mathematical theorems giving both necessary and sufficient conditions for
positivity (the \textquotedblleft KLM\ conditions\textquotedblright, which we
shortly review in Section \ref{sec2}), these results are not of great help in
practice because they involve the simultaneous verification of the positivity
of \textit{infinitely} many square matrices of increasing dimension. A
supplementary difficulty is actually lurking in the shadows: these conditions
are sensitive to the value of Planck's constant when the latter is used as a
variable parameter: a given operator $\widehat{\rho}$ might thus very well be
positive for one value of $\hbar$ and negative for another (this somewhat
unexpected but crucial property is best understood in terms of the
Narcowich--Wigner spectrum \cite{na88,brwe95}). This feature is of course
completely fatal when one wants to use semiclassical or WKB methods. In fact,
in a recent very interesting Letter \cite{mamamasuza06} Man'ko \textit{et al}.
have shown that rescaling the position and momentum coordinates by a common
factor can take a density matrix into a non-positive operator while preserving
a class of sharp uncertainty relations (the Robertson--Schr\"{o}dinger
uncertainty principle, which we recall in Section \ref{sec2}). This allows
these authors to conclude that \textquotedblleft...\textit{the uncertainty
principle does not determine the quantum state}\textquotedblright. Man'ko
\textit{et al. }are of course right; in fact Narcowich and O'Connell
\cite{naoc86} had already shown in the mid 1980s that fulfilling the
uncertainty relations is necessary, but \textit{not sufficient}, to ensure the
positivity of $\widehat{\rho}$ (this example is described in next Section).

The goal of this Letter is threefold. First (Section \ref{sec2}), we
complement and explain from a somewhat different (and more critical)
perspective the results of Man'ko \textit{et al}. \cite{mamamasuza06} (Man'ko
\textit{et al}. claim that the uncertainty relations are \textit{necessary} to
ensure positivity: they are again right, of course, but they do not prove this
fundamental fact!). Secondly (Section \ref{sec3}) we propose a new criterion
for deciding when a phase-space function which is dominated at infinity by a
phase-space Gaussian is the Wigner distribution of a mixed (non necessarily
Gaussian!) state. Our approach is based on Hardy's uncertainty principle
\cite{ha33} for a function and its Fourier transform. Hardy's theorem goes
back to 1933: it is unfortunate that its usefulness in quantum mechanics has
apparently not been noticed before! We conclude our discussion in Section
\ref{secblob}) by linking our results to the notions of \textquotedblleft
quantum blob\textquotedblright\ and \textquotedblleft admissible
ellipsoid\textquotedblright\ introduced by the first author in
\cite{de03-1,de04-1,Birk}, and which provides a canonically invariant notion
of phase-space coarse-graining, which seems promising in various aspects of
phase-space quantization.\medskip

\noindent\textbf{Notation}. We work in $N$ degrees of freedom; the coordinates
of position vector $x$ are $x_{1},...,x_{N}$ and those of the momentum vector
$p$ are $p_{1},...,p_{N}$. Writing $x$ and $p$ as column vectors we set $z=%
\begin{pmatrix}
x\\
p
\end{pmatrix}
$. We denote by $\sigma(z,z^{\prime})$ the symplectic product: by definition
$\sigma(z,z^{\prime})=(z^{\prime})^{T}Jz=p\cdot x^{\prime}-p^{\prime}\cdot x$
where where $J=%
\begin{pmatrix}
0 & I\\
-I & 0
\end{pmatrix}
$ is the standard symplectic matrix. A $2N\times2N$ real matrix $S$ is
symplectic if and only if $S^{T}JS=SJS^{T}=J$.

\section{Canonical formulation of the uncertainty principle and positivity
\label{sec2}}

Let $\widehat{\rho}$ be a self-adjoint trace-class operator on $L^{2}%
(\mathbb{R}^{N})$; we have%
\begin{equation}
\widehat{\rho}\psi(x)=\int K(x,x^{\prime})\psi(x^{\prime})d^{N}x^{\prime}
\label{6}%
\end{equation}
where the kernel $K$ satisfies $K(x,x^{\prime})=\overline{K(x^{\prime},x)}$
and is square integrable on $\mathbb{R}^{N}\times\mathbb{R}^{N}$. The Wigner
distribution of $\widehat{\rho}$ is the real function%
\begin{equation}
W(z)=\left(  \tfrac{1}{2\pi\hbar}\right)  ^{N}\int e^{-\frac{i}{\hbar}p\cdot
y}K(x+\tfrac{1}{2}y,x-\tfrac{1}{2}y)d^{N}y\text{.} \label{7}%
\end{equation}
(see Littlejohn \cite{li86} for details; our choice of normalization is
consistent with that adopted in Weyl calculus, but it differs from that in
\cite{mamamasuza06} even in the case $\hbar=1$). We have%
\begin{equation}
\operatorname*{Tr}(\widehat{\rho})=\int K(x,x)d^{N}x=\int W(z)d^{2N}z\text{.}
\label{8}%
\end{equation}
Assume now that in addition $\operatorname*{Tr}(\widehat{\rho})=1$; if
$\widehat{\rho}$ is positive then it is called a \textit{density matrix}. If
$A$ and $B$ are two essentially self-adjoint operators defined on some common
dense subset of $L^{2}(\mathbb{R}^{N})$ the \textit{covariance} of the pair
$(A,B)$ with respect to $\widehat{\rho}$ is by definition
\begin{equation}
\Delta(A,B)_{\widehat{\rho}}=\tfrac{1}{2}\left\langle AB+BA\right\rangle
_{\widehat{\rho}}-\left\langle A\right\rangle _{\widehat{\rho}}\left\langle
B\right\rangle _{\widehat{\rho}} \label{9}%
\end{equation}
where $\left\langle A\right\rangle _{\widehat{\rho}}=\operatorname*{Tr}%
(A\widehat{\rho})$, and so on. Choosing in particular for $A$ the position
operator $X_{j}=x_{j}$ and for $B$ the momentum operator $P_{j}=-i\hbar
\partial/\partial x_{j}$ the \textit{covariance matrix} is the symmetric
$2N\times2N $ matrix%
\begin{equation}
\Sigma_{\widehat{\rho}}=%
\begin{pmatrix}
\Delta(X,X)_{\widehat{\rho}} & \Delta(X,P)_{\widehat{\rho}}\\
\Delta(P,X)_{\widehat{\rho}} & \Delta(P,P)_{\widehat{\rho}}%
\end{pmatrix}
\label{10}%
\end{equation}
where $\Delta(X,X)_{\widehat{\rho}}=(\Delta(X_{j},X_{k})_{\widehat{\rho}%
})_{1\leq j,k\leq N}$, $\Delta(X,P)_{\widehat{\rho}}=(\Delta(X_{j}%
,P_{k})_{\widehat{\rho}})_{1\leq j,k\leq N}$ and so on (we assume that all
second moments exist; this condition is satisfied for instance if
$(1+|z|^{2})\rho$ is absolutely integrable). The covariance matrix is a
fundamental object in both classical and quantum statistical mechanics because
it incorporates the correlations between the considered variables. The
Robertson--Schr\"{o}dinger uncertainty principle says that
\begin{align}
(\Delta X_{j})_{\widehat{\rho}}^{2}(\Delta P_{j})_{\widehat{\rho}}^{2}  &
\geq\Delta(X_{j},P_{j})_{\widehat{\rho}}^{2}+\tfrac{1}{4}\hbar^{2}\text{ ,
}1\leq j\leq N\label{ua}\\
(\Delta X_{j})_{\widehat{\rho}}^{2}(\Delta P_{k})_{\widehat{\rho}}^{2}  &
\geq\Delta(X_{j},P_{k})_{\widehat{\rho}}^{2}\text{ \ for \ }j\neq k.
\label{ub}%
\end{align}
where $\Delta(X,P)$ is the covariance of the pair $(X_{j},P_{j})$ (see the
original articles \cite{bob34,schr} and the historical discussion by Trifonov
and Donev \cite{trdo98}; for a \textquotedblleft modern\textquotedblright%
\ proof in the general case of non-commuting observables the reader could
consult Messiah's classical treatise \cite{Messiah}).

One has the following fundamental result well-known in quantum optics, and
used in the study of entanglement and separability (see for instance
\cite{sisumu87,simudu96}):

\begin{quotation}
(I) \textit{If the self-adjoint trace-class operator} $\widehat{\rho}$ with
$Tr(\widehat{\rho})=1$ \textit{is positive, that is if it is a density matrix,
then the Hermitian matrix} $\Sigma_{\widehat{\rho}}+\frac{i\hbar}{2}J$
\textit{is positive semi-definite}:%
\begin{equation}
\Sigma_{\widehat{\rho}}+\frac{i\hbar}{2}J\geq0. \label{uc}%
\end{equation}

\end{quotation}

\noindent(That $\Sigma_{\widehat{\rho}}+\frac{i\hbar}{2}J$ is Hermitian
results from the symmetry of $\Sigma_{\widehat{\rho}}$ and the fact that
$J^{T}=-J$). The relation between property (I) and the
Robertson--Schr\"{o}dinger uncertainty principle is the following:

\begin{quotation}
(II) \textit{Condition (\ref{uc}) is equivalent to the
Robertson--Schr\"{o}dinger inequalities (\ref{ua})--(\ref{ub}).}
\end{quotation}

That the formulation (\ref{uc}) of the uncertainty principle is invariant
under linear canonical transformations follows at once from the fact that $S$
is a symplectic matrix; this makes the superiority of this formulation on the
usual one: it replaces the quite complicated and tedious verification of the
inequalities (\ref{ua})--(\ref{ub}) by the calculation of a set of
eigenvalues. To see why, let us begin by giving two definitions. Let $M$ be
any real positive-definite $2N\times2N$ matrix. Since $JM$ is equivalent to
the antisymmetric matrix $M^{1/2}JM^{1/2}$ its eigenvalues are of the type
$\pm i\mu_{j}$ ($j=1,...,N$) with $\mu_{j}>0$. Ordering the $\mu_{j}$ so that
$\mu_{1}\geq\mu_{2}\geq\cdots\geq\mu_{N}$ we call the sequence $(\mu
_{1},...,\mu_{N})$ the \emph{symplectic spectrum} of $M$; the number $\mu
=\mu_{1}$ is called the \emph{Williamson invariant} of $M$. Williamson
\cite{wi63} has proved that that there exists a $2N\times2N$ symplectic matrix
such that $M=S^{T}DS$ with%
\begin{equation}
D=%
\begin{pmatrix}
\Lambda & 0\\
0 & \Lambda
\end{pmatrix}
\text{ \ , \ }\Lambda=\operatorname*{diag}(\mu_{1},...,\mu_{N}) \label{de}%
\end{equation}
(\textquotedblleft Williamson diagonal form\textquotedblright). Now, one
proves \cite{sisumu87,simudu96} (see \cite{Birk} for a detailed exposition),
that condition (\ref{uc}) (and hence the Robertson--Schr\"{o}dinger
inequalities) is equivalent to:

\begin{quotation}
(III)\ \textit{The Williamson invariant }$\mu$ \textit{of the covariance
matrix }$\Sigma_{\widehat{\rho}}$ \textit{satisfies} $\mu\geq\frac{1}{2}\hbar$.
\end{quotation}

We emphasize that the equivalent conditions (I)--(III) are not sufficient to
ensure positivity; an illustration is the example of Narcowich and O'Connell
\cite{naoc86} mentioned in the Introduction. It goes as follows: let the
function $W(x,p)$ be determined by its Fourier transform via%
\begin{equation}
\int e^{i(xx^{\prime}+pp^{\prime})}W(x^{\prime},p^{\prime})dp^{\prime
}dx^{\prime}=(1-\tfrac{1}{2}\alpha x^{2}-\tfrac{1}{2}\beta p^{2}%
)e^{-(\alpha^{2}x^{4}+\beta^{2}p^{4})} \label{ass}%
\end{equation}
(with $\alpha,\beta>0$). It is easily checked that $W$ is real and that the
corresponding operator $\widehat{\rho}$ satisfies $\operatorname*{Tr}%
(\widehat{\rho})=1$. Narcowich and O'Connell then show that the uncertainty
principle is satisfied as soon as $\alpha$ and $\beta$ are chosen such that
$\alpha\beta\geq\hbar^{2}/4$. However, even with that choice, the operator
$\widehat{\rho}$ is never non-negative because the average of $p^{4}$ is in
all cases given by
\begin{equation}
\int p^{4}W(x,p)dxdp=-24\alpha^{2}<0\text{;} \label{5}%
\end{equation}
$\rho$ can thus not be the density matrix of any quantum state.

The considerations above explain the difficulties with positivity questions
occurring when one rescales Wigner distributions as Man'ko \textit{et al }do
in\textit{\ }\cite{mamamasuza06}. Let in fact $z_{\alpha}$ denote any of the
components of the vector $z=(x_{1},...,x_{N};p_{1},...,p_{N})$ and let
$Z_{\alpha},Z_{\beta}$ be the operators corresponding to $z_{\alpha},z_{\beta
}$. Defining the rescaled Wigner distribution $W^{\lambda}$ by%
\begin{equation}
W^{\lambda}(z)=\lambda^{-2N}W(\lambda z)\text{ \ , \ }\lambda>0 \label{11}%
\end{equation}
we have
\begin{equation}
\int W^{\lambda}(z)d^{2N}z=\int W(z)d^{2N}z=1 \label{12}%
\end{equation}
and a straightforward calculation shows that%
\begin{equation}
\Delta(Z_{\alpha},Z_{\beta})_{\widehat{\rho}^{\lambda}}=\frac{1}{\lambda^{2}%
}\Delta(Z_{\alpha},Z_{\beta})_{\widehat{\rho}} \label{13}%
\end{equation}
where $\widehat{\rho}^{\lambda}$ is the operator corresponding to $W^{\lambda
},$ hence
\begin{equation}
\Sigma_{\widehat{\rho}^{\lambda}}=\frac{1}{\lambda^{2}}\Sigma_{\widehat{\rho}%
}. \label{14}%
\end{equation}
Let $\mu$ and $\mu^{\lambda}$ be the Williamson invariants of the matrices
$\Sigma_{\widehat{\rho}}$ and $\Sigma_{\widehat{\rho}^{\lambda}}$,
respectively. In view of condition (III) in last section, we must have
$\mu\geq\frac{1}{2}\hbar.$ If now $\widehat{\rho}^{\lambda}$ is also to be a
density operator we must have $\mu^{\lambda}\geq\frac{1}{2}\hbar$ as well,
that is, equivalently $\mu\geq\frac{1}{2}\lambda^{-2}\hbar$. This requires
that $\lambda\leq1$.

In the Introduction section of this Letter we referred to necessary and
sufficient conditions for a self-adjoint trace-class operator to be positive.
In fact, Kastler \cite{ka65} and Loupias and Miracle-Sole
\cite{LouMiracle1,LouMiracle2} have shown that the operator $\widehat{\rho}$
is positive (and hence a density operator) if and only the following so-called
\textquotedblleft KLM\ conditions\textquotedblright\ hold: for every integer
$m=1,2,...$ the complex matrix $F=(F_{jk}(z_{j},z_{k}))_{1\leq j,k\leq m}$
with
\begin{equation}
F_{jk}(z_{j},z_{k})=e^{\frac{i\hbar}{2}\sigma(z_{j},z_{k})}\mathcal{F}%
_{\sigma}W(z_{j}-z_{k}) \label{15}%
\end{equation}
is positive semi-definite; here
\begin{equation}
\mathcal{F}_{\sigma}W_{\rho}(z)=\int e^{i\sigma(z,z^{\prime})}W(z^{\prime
})d^{N}z^{\prime} \label{16}%
\end{equation}
is the symplectic Fourier transform of the Wigner distribution $\rho$.

In \cite{na90}, Lemma 2.1, Narcowich shows that the KLM conditions imply that
$\Sigma_{\widehat{\rho}}+\frac{i\hbar}{2}J\geq0$. In view of (I), (II) above
we thus have the following \textit{necessary} condition for $\widehat{\rho}$
to be positive:

\begin{quotation}
(IV) \textit{Assume that the operator} $\widehat{\rho}$ \textit{is positive;
then its covariance matrix must satisfy condition (\ref{uc}) or, equivalently,
the Schr\"{o}dinger--Robertson inequalities (\ref{ua})--(\ref{ub})}.
\end{quotation}

This statement thus fully justifies and completes the statement of Man'ko et
al. \cite{mamamasuza06} that \textquotedblleft...\textit{the uncertainty
principle does not determine the quantum state}\textquotedblright.

\section{Gaussian estimates and Hardy's theorem\label{sec3}}

We ask the following question:

\begin{quotation}
\textquotedblleft\textit{Under which conditions on }$M$\textit{\ can a
function }$W$\textit{\ such that }$W(z)\leq Ce^{-\frac{1}{\hbar}Mz\cdot z}$
\textit{be the Wigner distribution of some mixed sate?}\textquotedblright.
\end{quotation}

The answer to that question is given by following theorem, the proof of which
relies on the following old result due to Hardy \cite{ha33}: assume that the
square-integrable function $\psi$ and its Fourier transform
\[
F\psi(p)=\left(  \tfrac{1}{2\pi\hbar}\right)  ^{N}\int e^{-\frac{i}{\hbar
}p\cdot x}\psi(x)d^{N}x
\]
are such that $|\psi(x)|\leq Ce^{-\frac{a}{2\hbar}|x|^{2}}$ and $|F\psi
(p)|\leq Ce^{-\frac{b}{2\hbar}|p|^{2}}$ ($a$, $b>0$). Then we must have
$ab\leq1,$ and if $ab=1$ then $\psi(x)=Ae^{-\frac{a}{2\hbar}|x|^{2}}$ for some
complex constant $A$. If $ab>1$ then $\psi=0$.

\begin{theorem}
\label{un}Let $\widehat{\rho}$ be a density operator and assume that its
Wigner distribution $W$ satisfies an estimate%
\begin{equation}
W(z)\leq Ce^{-\frac{1}{\hbar}Mz\cdot z} \label{est}%
\end{equation}
where $C>0$ is some constant and $M=M^{T}>0$. Then the Williamson invariant
$\mu$ of $M$ must satisfy $\mu_{1}\leq1$. Equivalently: the matrix
$\Sigma_{\widehat{\rho}}=\frac{\hbar}{2}M^{-1}$ must satisfy the uncertainty
principle: $\Sigma_{\widehat{\rho}}+\frac{i\hbar}{2}J\geq0$.
\end{theorem}

To prove this we will use the fact that there exists an orthonormal system of
vectors $(\psi_{j})_{j}$ in $L^{2}(\mathbb{R}^{N})$ such that
\begin{equation}
W(z)=\sum_{j}\alpha_{j}W\psi_{j}(z) \label{17}%
\end{equation}
with $\sum_{j}\alpha_{j}=1$, $\alpha_{j}>0$ (see e.g. \cite{Birk} and the
references therein). Let $S$ be a symplectic matrix such that $M=S^{T}DS$ with
$D$ as in (\ref{de}); then the inequality (\ref{est}) is equivalent to
\begin{equation}
W_{\rho}(S^{-1}z)\leq Ce^{-\frac{1}{\hbar}\Lambda x\cdot x}e^{-\frac{1}{\hbar
}\Lambda p\cdot p}. \label{18}%
\end{equation}
Integrating successively with respect to the variables $p_{j}$ and $x_{j}$ we
get%
\begin{equation}
\int W(S^{-1}z)dp\leq C_{\Lambda}e^{-\frac{1}{\hbar}\Lambda x\cdot x}\text{
\ , \ }\int W(S^{-1}z)d^{N}x\leq C_{\Lambda}e^{-\frac{1}{\hbar}\Lambda p\cdot
p} \label{19}%
\end{equation}
with $C_{\Lambda}=C\int e^{-\frac{1}{\hbar}\Lambda x\cdot x}d^{N}x$. We next
observe that%
\begin{equation}
W(S^{-1}z)=\sum_{j}\alpha_{j}W\psi_{j}(S^{-1}z)=\sum_{j}\alpha_{j}%
W(\widehat{S}\psi_{j})(z) \label{20}%
\end{equation}
where $\widehat{S}$ is any of the two metaplectic operators corresponding to
$S$ (see for instance \cite{Birk,li86} and the references therein). Taking
into account the formulae%
\begin{equation}
\int W(\widehat{S}\psi_{j})(z)d^{N}p=|\widehat{S}\psi_{j}(x)|^{2}\text{ ,
}\int W(\widehat{S}\psi_{j})(z)d^{N}x=|F(\widehat{S}\psi_{j})(p)|^{2}
\label{21}%
\end{equation}
it follows that we have%
\begin{equation}
\sum_{j}\alpha_{j}|\widehat{S}\psi_{j}(x)|^{2}\leq C_{\Lambda}e^{-\frac
{1}{\hbar}\Lambda x\cdot x}\text{ \ , \ }\sum_{j}\alpha_{j}|F(\widehat{S}%
\psi_{j})(p)|^{2}\leq C_{\Lambda}e^{-\frac{1}{\hbar}\Lambda p\cdot p}
\label{22}%
\end{equation}
and hence, in particular,
\begin{equation}
|\widehat{S}\psi_{j}(x)|\leq C_{j,\Lambda}e^{-\frac{1}{2\hbar}\Lambda x\cdot
x}\text{ \ , \ }|F(\widehat{S}\psi_{j})(p)|\leq C_{j,\Lambda}e^{-\frac
{1}{2\hbar}\Lambda p\cdot p} \label{23}%
\end{equation}
with $C_{j,\Lambda}=\sqrt{C_{\Lambda}/\alpha_{j}}$. Since $\Lambda
=\operatorname*{diag}(\mu_{1},...,\mu_{N})$ with $\mu_{1}\geq\cdots\geq\mu
_{N}$ it follows that
\begin{equation}
|\widehat{S}\psi_{j}(x)|\leq C_{j,\Lambda}e^{-\frac{1}{2\hbar}\Lambda x\cdot
x}\text{ \ , \ }|F(\widehat{S}\psi_{j})(p)|\leq C_{j,\Lambda}e^{-\frac
{1}{2\hbar}\Lambda p\cdot p}. \label{ineq}%
\end{equation}
We claim that these inequalities can only hold if $\mu_{1}\leq1$. Set
\[
\phi_{j}(x_{1})=\widehat{S}\psi_{j}(x_{1},0,...,0).
\]
By the first inequality (\ref{ineq}) we have
\begin{equation}
|\phi_{j}(x_{1})|\leq C_{j,\Lambda}e^{-\frac{\mu_{1}}{2\hbar}x_{1}^{2}}.
\label{trois}%
\end{equation}
Denoting by $F_{1}$ the Fourier transform in the variable $x_{1}$ a
straightforward calculation shows that
\begin{equation}
\int F(\widehat{S}\psi_{j})(p)dp_{2}\cdot\cdot\cdot dp_{N}=\left(  2\pi
\hbar\right)  ^{(N-1)/2}F_{1}\phi_{j}(p_{1}). \label{24}%
\end{equation}
and hence, in view of the second inequality (\ref{ineq}),
\begin{equation}
|F_{1}\phi_{j}(p_{1})|\leq\left(  \tfrac{1}{2\pi\hbar}\right)  ^{(N-1)/2}%
C_{j,\Lambda}\int e^{-\frac{1}{2\hbar}\sum_{j=1}^{N}\mu_{j}p_{j}^{2}}%
dp_{2}\cdot\cdot\cdot dp_{N} \label{25}%
\end{equation}
that is%
\begin{equation}
|F_{1}\phi_{j}(p_{1})(p_{1})|\leq C_{j,\Lambda}e^{-\frac{\mu_{1}}{2\hbar}%
p_{1}^{2}} \label{quatre}%
\end{equation}
for a new constant $C_{j,\Lambda}$. Applying Hardy's theorem to (\ref{trois})
and (\ref{quatre}) we must have $\mu_{1}^{2}\leq1$, which proves our claim.

The result above shows the reason for which the rescaling can be used to
produce negative operators from a positive one: assume that $W_{\rho}(z)\leq
Ce^{-\frac{1}{\hbar}Mz\cdot z}$; then
\begin{equation}
\lambda^{-2N}W_{\rho}(\lambda z)\leq\lambda^{-2N}C_{\lambda}e^{-\frac{1}%
{\hbar}M_{\lambda}z\cdot z} \label{26}%
\end{equation}
with $C_{\lambda}=\lambda^{-2N}C$ and $M_{\lambda}=\lambda^{2}M$. The
Williamson invariant of $M_{\lambda}$ is $\lambda^{2}\mu_{1}$ and the
condition $\lambda^{2}\mu_{1}\leq1$ will be violated as soon as we choose
$\lambda>1/\sqrt{\mu_{1}}$. (In \cite{mamamasuza06} Man'ko \textit{et al}.
work in units in which $\hbar=1$; it is therefore not immediately obvious that
the procedure they implement to construct non-positive operators by rescaling
coordinates in a non-symplectic way is tantamount to increasing the value of
Planck's constant so that the uncertainty principle is violated.) In fact, one
immediately understands why the rescaling procedure of Man'ko \textit{et al.}
works: it consists (in the example they consider) in replacing $M$ by a matrix
that is \textquotedblleft too small\textquotedblright. In addition, as a
by-product of our result we recover the property that the support of a Wigner
distribution can never be bounded in phase space.

\begin{remark}
Theorem \ref{un} also allows us to recover in a simple way a result of Folland
and Sitaram \cite{fosi97}: \textit{a Wigner distribution can never be
compactly supported}. Suppose indeed that there exists some $R>0$ such that
$W_{\rho}(z)=0$ for $|z|\geq R$. For any given $\mu$ we can always choose
$C>0$ large enough so that $W_{\rho}(z)\leq Ce^{-\frac{\mu}{\hbar}|z|^{2}}$
for all $z$. Choosing $\mu>1$ this contradicts the theorem..
\end{remark}

\section{Relation with Quantum Blobs\label{secblob}}

In recent previous work \cite{de03-1,de04-1} one of us has introduced the
notion of \textquotedblleft quantum blob\textquotedblright\ and of
\textquotedblleft admissible ellipsoid\textquotedblright\ in connection with
the study of a coordinate-free formulation of the uncertainty principle. A
quantum blob is the image of a phase-space ball with radius $\sqrt{\hbar}$ by
a (linear or affine) symplectic translation. An admissible ellipsoid is a
phase-space ellipsoid containing a quantum blob. Characteristic properties are:

\begin{itemize}
\item The section of a quantum blob by any plane through its center which is
parallel to a plane of conjugate coordinates $x_{j},p_{j}$ has area $\frac
{1}{2}h$;

\item A phase-space ellipsoid is admissible if and only if its section by any
plane through its center which is parallel to a plane of conjugate coordinates
$x_{j},p_{j}$ has area at least $\frac{1}{2}h$.
\end{itemize}

Moreover:

\begin{quotation}
(V) \textit{An ellipsoid }$\mathcal{B}_{M}:Mz\cdot z\leq\hbar$ is admissible
if and only if $c(\mathcal{B}_{M})\geq\frac{1}{2}h$, $c$ any symplectic
capacity \cite{hoze94} on $\mathbb{R}^{2N}$, and this condition is equivalent
to $\Sigma+\frac{i\hbar}{2}J\geq0$ with $\Sigma=\frac{\hbar}{2}M^{-1}$.
\end{quotation}

and

\begin{quotation}
(VI) \textit{The symplectic capacity of} $\mathcal{B}_{M}:Mz\cdot z\leq\hbar$
is $c(\mathcal{B}_{M})=\pi\hbar/\mu_{1}$ where $(\mu_{1},...,\mu_{N})$ is the
symplectic spectrum of $\mathcal{B}_{M}$ (see \cite{Birk,hoze94}).
\end{quotation}

We can thus re-express Theorem \ref{un} in the following coordinate-free form:

\begin{theorem}
\label{deux}Assume that the Wigner distribution of a density operator
$\widehat{\rho}$ is such that $W(z)\leq Ce^{-\frac{1}{\hbar}Mz\cdot z}$. Then
$c(\mathcal{B}_{M})\geq\frac{1}{2}h$ where $\mathcal{B}_{M}$ is the ellipsoid
$Mz\cdot z\leq\hbar$
\end{theorem}

It can be interpreted in a very visual way as follows: assume that we have
coarse-grained phase space by quantum blobs $S(B(\sqrt{\hbar}))$. Then the
Wigner ellipsoid of a density operator cannot be arbitrarily small, but must
contain such a quantum blob. Equivalently: the Wigner ellipsoid must be
defined on the \textquotedblleft quantum phase-space\textquotedblright%
\ consisting of all parts of $\mathbb{R}^{2N}$ containing a quantum blob.

\section{Conclusion and Comments}

A \textquotedblleft simple\textquotedblright\ characterization of positivity
for trace-class operators is still to be found. We have given one such
characterization for a particular class of putative Wigner distributions
(those dominated by a phase-space Gaussian). In the general case possibly the
phase-space techniques and concepts (symplectic capacities) developed in
\cite{de03-1,de04-1} could provide further insight about what such a condition
could be (cf. Theorem \ref{deux} above). The methods proposed in Bohm and
Hiley \cite{bohi81} could perhaps shed some light on the question; also see
Bracken and Wood \cite{brwo04} who introduce the interesting notion of
\textquotedblleft Groenewold operator\textquotedblright\ to study positivity
(but from a slightly different point of view).

We finally remark that in the discussed paper \cite{mamamasuza06} Man'ko
\textit{et al.} use the notion of quantum fidelity (which, besides, plays an
important role in the study of Loschmidt echo) to prove that the Wigner
function of the first excited state of the oscillator does not lead to a
positive operator when rescaled; perhaps their idea could be exploited in a
more general context to shed some light on the difficult question of
positivity? We will come back to these fundamental questions in a forthcoming
paper.\bigskip

\noindent\textbf{Acknowledgements}. \textit{Both authors have been financed by
the European Union Marie Curie Excellence grant MEXT-CT-2004-517154.}


\begin{thebibliography}{99}                                                                                               %
\bibitem {na88}F.J. Narcowich, J.\ Math. Phys. 29(9) (1988) 2036--2041.

\bibitem {brwe95}T. Br\"{o}cker and R.F. Werner. J.\ Math. Phys. 36(1) (1990) 62--75.

\bibitem {mamamasuza06}O.V. Man'ko, V.I. Man'ko, G. Marmo, E.C.G. Sudarshan,
and F. Zaccaria, Phys. Lett. A 357 (2006) 255--260.

\bibitem {naoc86}F.J. Narcowich and R.F. O'Connell, Phys. Rev. A 34(1) (1986) 1--6.

\bibitem {ha33}G.H.\ Hardy. J. London. Math. Soc. 8 (1933) 227--231.

\bibitem {de03-1}M. de Gosson. Phys. Lett. A, 317/5-6 (2003) 365--369.

\bibitem {de04-1}M. de Gosson. Phys. Lett. A, 330:3--4 (2004) 161--167.

\bibitem {Birk}M. de Gosson. Symplectic Geometry and Quantum Mechanics.
Birkh\"{a}user, Basel, series \textquotedblleft Operator Theory: Advances and
Applications\textquotedblright\ (subseries: \textquotedblleft Advances in
Partial Differential Equations\textquotedblright), Vol. 166 (2006).

\bibitem {li86}R.G. Littlejohn. Phys. Reports 138(4--5) (1986) 193--291.

\bibitem {bob34}H.P. Robertson. Phys. Rev. 34 (1929) 163.

\bibitem {schr}E. Schr\"{o}dinger. Sitz. der Preuss. Acad. Wiss. (Phys. Math.
Klasse) (1930) 296.

\bibitem {trdo98}D.A. Trifonov and S.G. Donev. J. Phys. A: Math. Gen. (1998) 8041--8047.

\bibitem {Messiah}A. Messiah. M\'{e}canique Quantique. Dunod, Paris, 1961,
1995 (Vol. 1) [English translation: Quantum Mechanics, North--Holland, 1991].

\bibitem {sisumu87}R. Simon, E.C.G. Sudarshan, and N. Mukunda. Phys. Rev. A
36(8) (1987) 3868--3880.

\bibitem {simudu96}R. Simon, N. Mukunda, and B. Dutta. Phys. Rev. A 49 (1994) 1567--1583.

\bibitem {na90}F.J. Narcowich, J.\ Math. Phys. 31(2) (1990) 354--364.

\bibitem {wi63}J. Williamson. Amer. J. of Math. 58 (1936) 141--163.

\bibitem {ka65}D. Kastler. Commun. math. Phys. 1 (1965) 14--48.

\bibitem {LouMiracle1}G. Loupias et S. Miracle-Sole. Commun. math. Phys. 2
(1966), 31--48.

\bibitem {LouMiracle2}G. Loupias et S. Miracle-Sole. Ann. Inst. Henri
Poincar\'{e} 6(1) (1967) 39--58.

\bibitem {hoze94}H. Hofer et E. Zehnder. Symplectic Invariants and Hamiltonian
Dynamics. Birkh\"{a}user Advanced texts, Birkh\"{a}user Verlag, 1994.

\bibitem {bohi81}D. Bohm and B. Hiley. Foundations of Physics, 11(3/4) (1981) 179--206.

\bibitem {brwo04}A.J. Bracken and J.G. Wood. Europhys. Lett. 68(1) (2004) 1--7.

\bibitem {hiocscwi84}M. Hillery, R.F. O'Connell, M.O. Scully, and E.P. Wigner.
Phys. Reports 106(3) (1984) 121--167.

\bibitem {fosi97}G.B. Folland et A. Sitaram. Journ. Fourier Anal. Appl. 3(3)
(1997) 207--238.
\end{thebibliography}
\end{document}